\pdfoutput=1
\documentclass[aps,prl,twocolumn,superscriptaddress]{revtex4-1}
\usepackage{calrsfs}
\usepackage{mathrsfs}
\usepackage{graphicx}
\usepackage{dcolumn}
\usepackage{bm}

\usepackage{hyperref}
\usepackage{xcolor}
\hypersetup{colorlinks=true,linkcolor=red}
\usepackage{color}
\usepackage{graphicx}
\usepackage{fancyhdr}
\usepackage{url}
\usepackage{amssymb,amsfonts,amsmath,amsbsy,bm,t1enc,lipsum,latexsym}

\begin{document}

\preprint{APS/123-QED}

\title{Frequency Comb Assisted Broadband Precision Spectroscopy \\ 
with Cascaded Diode Lasers}


\author{Junqiu Liu}
\affiliation{{\'E}cole Polytechnique F{\'e}d{\'e}rale de Lausanne (EPFL), CH-1015 Lausanne, Switzerland}
\affiliation{Erlangen Graduate School in Advanced Optical Technologies, Friedrich-Alexander-Universit\"{a}t Erlangen-N\"{u}rnberg, 91052 Erlangen, Germany}

\author{Victor Brasch}
\affiliation{{\'E}cole Polytechnique F{\'e}d{\'e}rale de Lausanne (EPFL), CH-1015 Lausanne, Switzerland}

\author{Martin H. P. Pfeiffer}
\affiliation{{\'E}cole Polytechnique F{\'e}d{\'e}rale de Lausanne (EPFL), CH-1015 Lausanne, Switzerland}

\author{Arne Kordts}
\affiliation{{\'E}cole Polytechnique F{\'e}d{\'e}rale de Lausanne (EPFL), CH-1015 Lausanne, Switzerland}

\author{Ayman N. Kamel}
\affiliation{{\'E}cole Polytechnique F{\'e}d{\'e}rale de Lausanne (EPFL), CH-1015 Lausanne, Switzerland}
\affiliation{Present addresss: DTU Fotonik, Department of Photonics Engineering, Technical University of Denmark, DK-2800 Lyngby, Denmark}

\author{Hairun Guo}
\affiliation{{\'E}cole Polytechnique F{\'e}d{\'e}rale de Lausanne (EPFL), CH-1015 Lausanne, Switzerland}

\author{Michael Geiselmann}
\affiliation{{\'E}cole Polytechnique F{\'e}d{\'e}rale de Lausanne (EPFL), CH-1015 Lausanne, Switzerland}

\author{Tobias J. Kippenberg}
\email[E-mail: ]{tobias.kippenberg@epfl.ch}
\affiliation{{\'E}cole Polytechnique F{\'e}d{\'e}rale de Lausanne (EPFL), CH-1015 Lausanne, Switzerland}



\begin{abstract}
  Frequency comb assisted diode laser spectroscopy, employing both the accuracy of an optical frequency comb and the broad wavelength tuning range of a tunable diode laser, has been widely used in many applications. In this letter we present a novel method using cascaded frequency agile diode lasers, which allows extending the measurement bandwidth to 37.4 THz (1355 -- 1630 nm) at MHz resolution with scanning speeds above 1 THz/s. It is demonstrated as a useful tool to characterize a broadband spectrum for molecular spectroscopy and in particular it enables to characterize the dispersion of integrated microresonators up to the fourth order.


\end{abstract}

\keywords{Spectroscopy, diode lasers; Dispersion; Microcavities.}

\maketitle

Frequency combs \cite{Udem:02, Cundiff:03}, providing an equidistant grid of lines with precisely known frequencies over a broad spectral range, have substantially advanced precision spectroscopy over the past decades. To date, diverse spectroscopic methods employing frequency combs have been invented, such as direct frequency comb spectroscopy\cite{Foltynowicz:11}, Fourier transform spectroscopy \cite{Mandon:09} and dual-comb spectroscopy \cite{Bernhardt:10}. Among these methods, frequency comb assisted diode laser spectroscopy \cite{DelHaye:09}, enabling broadband spectral characterization with fast measurement speed (> 1 THz/s) and simple implementation, has been successfully applied for distance measurement \cite{Baumann:13, Baumann:14}, dynamic waveform detection \cite{Giorgetta:10}, plasma diagnostics \cite{Urabe:12} and molecular spectroscopy \cite{Nishiyama:13, Nishiyama:14}. One application benefiting from these advantages is the dispersion characterization of high-Q microresonators \cite{Riemensberger:12, Herr:14, Kordts:16, DelHaye:15, Pfeiffer:16}, while alternative methods using direct frequency comb \cite{Thorpe:05, Schliesser:06}, white light source \cite{Savchenkov:08} or sideband spectroscopy \cite{Li:12} have several limitations including system complexity, low measurement speed, narrow bandwidth and inability to measure microresonators with free spectral ranges (FSR) exceeding 100 GHz.



\begin{figure}[t]
\centering
\includegraphics[width=1 \linewidth]{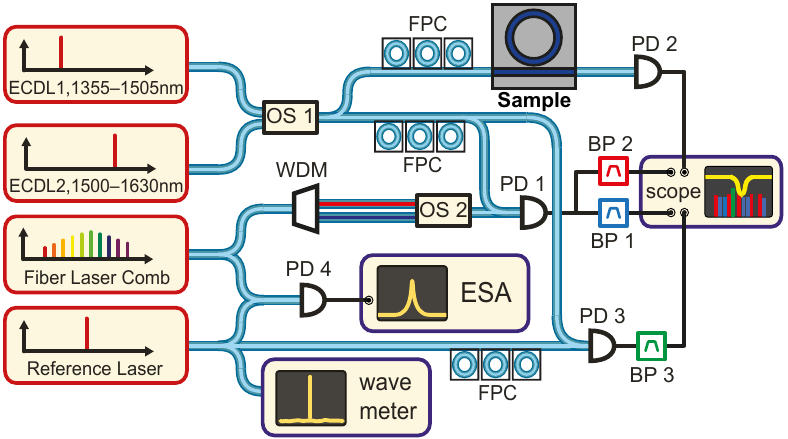}
\caption{Schematic of the setup used for cascaded frequency comb assisted diode laser spectroscopy. ECDL, external cavity diode laser; OS, optical switch; FPC, fiber polarization controller; WDM, wavelength-division multiplexer; BP, bandpass filter; PD, photodiode; ESA, electrical spectrum analyzer. }
\label{Fig:figure1}
\end{figure}

The dispersion characterization is important for the dispersion engineering of integrated high-Q microresonators for Kerr frequency comb generation \cite{DelHaye:07, Kippenberg:11} and bright dissipative Kerr soliton formation \cite{ HerrNP:14, Yi:15, Joshi:16, Brasch:15}. In addition, properly engineered higher order dispersion can lead to the emission of a dispersive wave via the process of soliton Cherenkov radiation \cite{Brasch:15, Milian:14, Jang:14, Karpov:16}. Several techniques based on geometry variation \cite{Yang:16} and additional material layers \cite{Riemensberger:12, Jiang:14} have been demonstrated to tailor the dispersion. However, to measure the higher order dispersion of microresonators, frequency comb assisted diode laser spectroscopy is currently limited by its measurement bandwidth, which is mainly determined by the wavelength tuning range of the used laser. Therefore, using more than one laser to cover different spectral ranges is desired to overcome the bandwidth limitation thus enabling measuring the higher order dispersion. In this letter we demonstrate a method to extend the measurement bandwidth by cascading two widely tunable lasers covering the wavelength range from 1355 nm to 1630 nm. The validity of our method is examined by molecular absorption spectroscopy. We subsequently use this method to characterize the dispersion of a photonic chip-based silicon nitride (Si$_3$N$_4$) microresonator \cite{Moss:13} whose FSR is approximately 1 THz. This is the first time that higher order dispersion is directly measured for such microresonators.


The experimental setup is shown in Fig. \ref{Fig:figure1} and is based on the setup described in Ref. \citep{DelHaye:09}. Two widely tunable mode-hop-free external cavity diode lasers (ECDL 1, ECDL 2, Santec TSL-510) with wavelength tuning ranges of 1355 -- 1505 nm and 1500 -- 1630 nm beat with a fully-stabilized spectrally broadened erbium-doped-fiber-laser-based frequency comb (MenloSystems FC1500, repetition rate $f_\text{rep}\approx$250 MHz, range 1050 -- 2100 nm). A 1310 nm/1550 nm wavelength-division multiplexer (WDM) splits the frequency comb into two branches. Two optical switches are synchronized, such that ECDL 1 beats with the 1310 nm branch, and successively ECDL 2 beats with the 1550 nm branch. The purpose of using the WDM and synchronizing the optical switches is to suppress the comb lines which do not contribute to the beat signals, thus to prevent the photodiode saturation and to improve the signal-to-noise ratio of the beat signals.

\begin{figure}[t]
\centering
\includegraphics[width=1 \linewidth]{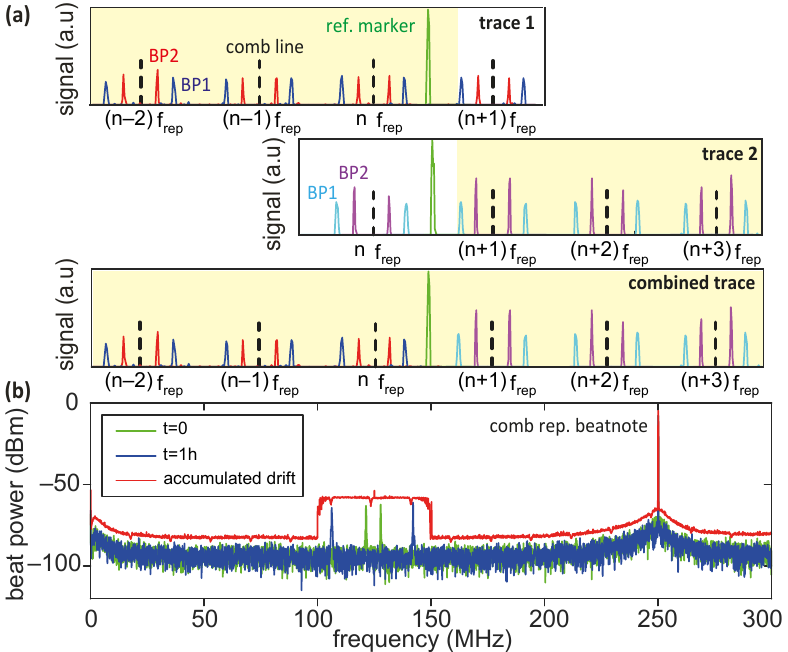}
\caption{Procedure to combine two ECDL scan traces using the reference laser. (a) The calibration markers (blue, red, cyan and purple) and the reference markers (green) are denoted in both traces. Two BPs are used in order to generate double calibration markers for better frequency determination. Each group of calibration markers is indexed by the comb line that generates it (black dashed lines). The combined trace is formed by combining the data in the yellow shaded zones of both traces. (b) Measurement of the reference laser's drift within 1 hour by beating it with the frequency comb and observing the beat note's drift on an ESA. The peak at 250 MHz is the frequency comb's repetition rate beat $f_\text{rep}$. Two peaks symmetrical to 125 MHz are due to the reference laser beating with its two nearest comb lines. Accumulated drift is recorded by using the ``maximum hold'' function of the ESA.}
\label{Fig:figure2}
\end{figure}

The two ECDLs scan one after the other with a scan speed of 10 nm/s. By using the band-pass filters labeled as BP 1, BP 2 (center frequencies $f_\text{BP 1}$, $f_\text{BP 2}$), the scanning ECDL generates four ``calibration markers'' per $f_\text{rep}$ interval when the frequency distance to its nearest comb line is $\pm f_\text{BP 1}$ or $\pm f_\text{BP 2}$ \cite{DelHaye:09}. The key problem is the combination of the two individual traces generated by the two ECDL scans into a single continuous trace which covers the full measurement range. This is solved by using an auxiliary reference laser whose wavelength $\lambda_\text{ref}$ is set in the 1500 -- 1505 nm range where both ECDLs overlap spectrally. With a low-pass filter (BP 3), in each trace a ``reference marker'' is recorded when the ECDL scans over the reference laser. The reference laser is set initially $f_\text{rep}/2$ from its nearest comb line, and its long-term drift is measured as <20 MHz/h as shown in Fig. \ref{Fig:figure2}(b). As long as the reference laser drifts less than $f_\text{rep}/2$ from its initial position within the measurement time ($\approx$60 s), which can be monitored by an electrical spectrum analyzer (ESA), we can unambiguously assume that the calibration markers adjacent to the reference marker are generated by the same comb line in both traces. Therefore, using the reference marker, the indices of the calibration markers in both traces can be matched. As shown in Fig. \ref{Fig:figure2}(a), by combining the data before the reference marker in trace 1 with the data after the reference marker in trace 2, a complete continuous trace from 1355 nm to 1630 nm is formed.

\begin{figure}[t]
\centering
\includegraphics[width=1 \linewidth]{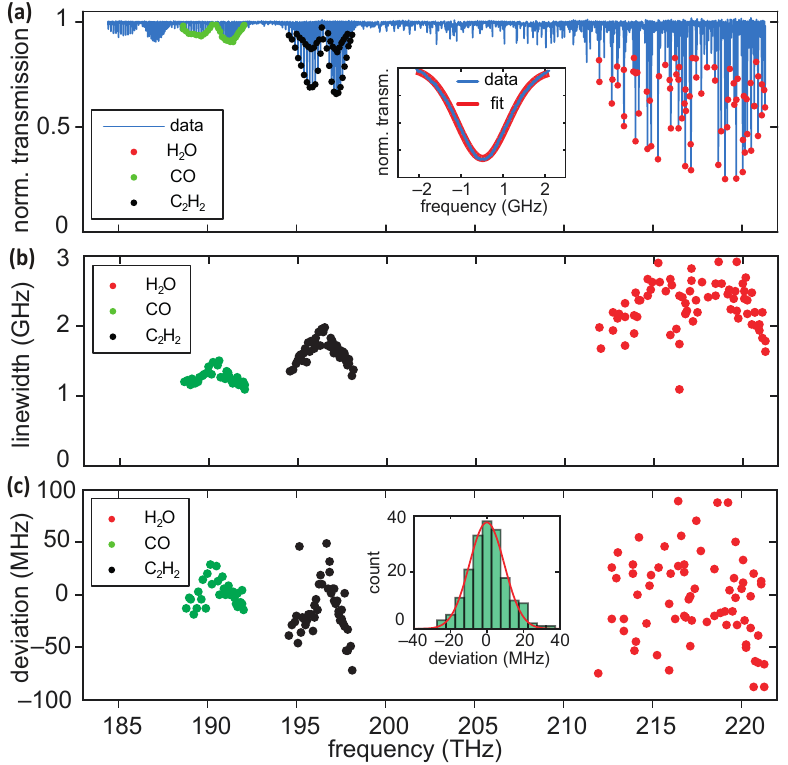}
\caption{Broadband characterization of a molecular absorption spectroscopy. (a) Normalized transmission spectrum of the gas cell. Each kind of molecule is marked according to the data from HITRAN\emph{online} \cite{Rothman:13}. Inset: one CO line and fit with Gaussian profiles. (b) Linewidths of each absorption line fitted with Gaussian profiles. (c) Deviations of the fitted line-center frequencies from the values on HITRAN\emph{online}. Inset: histogram of the deviations of CO lines from six repeated measurements and fit with a Gaussian distribution (red curve) with a standard deviation of 13.6 MHz.}
\label{Fig:figure3}
\end{figure}

Generally to combine individual traces into a single trace, it is required that: (1) each ECDL scans mode-hop-free; (2) any two adjacent ECDLs have a shared wavelength range; (3) a low-drift stable reference laser exists whose wavelength falls in each shared wavelength range; (4) calibration markers are well-resolved in each trace. Once the above conditions are satisfied, such setup can be extended with more than two ECDLs, enabling even broader measurement bandwidth.

\begin{figure}[t!]
\centering
\includegraphics[width=1 \linewidth]{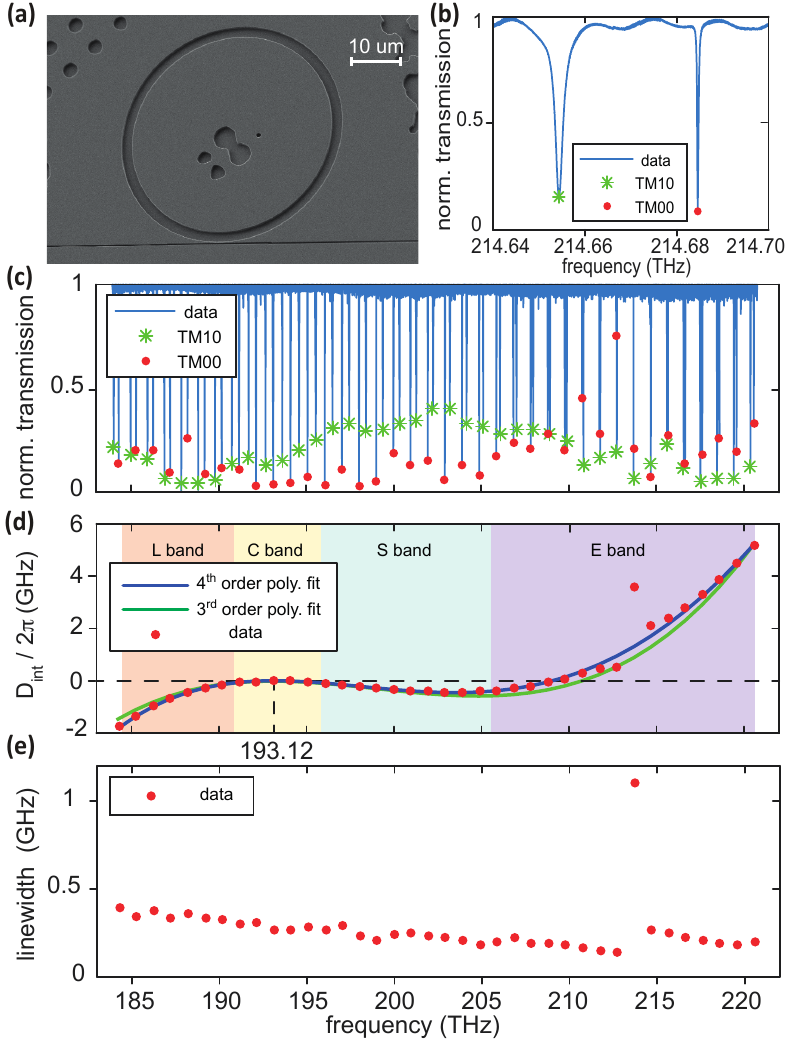}
\caption{Broadband dispersion characterization of a Si$_3$N$_4$ photonic chip-based microresonator with approximately 1 THz FSR. (a) SEM picture of the used microresonator geometry. (b) Linewidth difference between TM$_{00}$ and TM$_{10}$ modes. (c) Normalized transmission spectrum through the microresonator. (d) Plot of $D_\text{int}/2\pi$ of the TM$_{00}$ mode. Red points are the measured data, green solid curve is the 3$^\text{rd}$ order weighted fitting polynomial, blue solid curve is the 4$^\text{th}$ order weighted fitting polynomial. Different optical fiber communication bands are denoted by different color shaded zones. (e) Plot of the fitted linewidths of the TM$_{00}$ mode.}
\label{Fig:figure4}
\end{figure}

The reference laser sets the frequency reference point, and the frequency axis is calibrated with respect to the frequency comb based on $f_\text{rep}$. Assuming that the ECDLs scan uniformly in each $f_\text{rep}$ interval, the instantaneous frequency can be interpolated with respect to $f_\text{BP 1}$ and $f_\text{BP 2}$. The precision of frequency determination is limited by a 2.2 MHz/point resolution limit due to the oscilloscope's maximum 10 million data points per trace, and the error due to the 2 MHz bandwidth of the BPs. Therefore the total error is estimated as 4.2 MHz/point. However such error will not be prominent for spectral features with >100 MHz linewidth, as the center frequency is usually obtained by line profile fitting whose precision is then limited by the signal quality, fitting function used, resonance splitting etc.


The reference laser's wavelength $\lambda_\text{ref}$ is read from a wavelength meter with an imprecision of few picometers (corresponding frequency imprecision is few hundreds MHz). Such imprecision will introduce a global offset to the absolute frequency calibration, but will not affect the continuity of the trace combination and the relative frequency calibration with respect to $\lambda_\text{ref}$. The global offset (i.e. an absolute frequency measurement) is much less important for certain measurements such as the dispersion measurement of microresonators where the entire frequency scan needs to exhibit a precise relative frequency calibration, but not necessarily an absolute frequency calibration.


In order to examine the validity of our method, e.g. the continuity of the combined trace and the relative frequency calibration, we implemented a molecular absorption spectroscopy of a gas cell composed of water (H$_2$O), carbon monoxide (CO) and acetylene (C$_2$H$_2$), and compared it with the absorption line data from HITRAN\emph{online} \cite{Rothman:13} (www.hitran.org). The normalized transmission spectrum is shown in Fig. \ref{Fig:figure3}(a) and the absorption lines for each kind of molecule are marked. As the spectroscopy is not Doppler-free, a Gaussian function is used to fit the Doppler broadening profile \cite{Preston:96}, and the linewidth distribution is shown in Fig. \ref{Fig:figure3}(b). The fitted line-center frequencies $f_\text{fit}$ are then compared with the known frequencies $f_\text{HIT}$ from HITRAN\emph{online}, and a global offset $f_\text{off}\approx100$ MHz is observed due to the reason mentioned above. Subtracting the global offset, the frequency deviations defined by $\delta=f_\text{fit}-f_\text{HIT}-f_\text{off}$ are plotted in Fig. \ref{Fig:figure3}(c). All the frequency deviations are distributed in a $\pm$100 MHz range around 0 MHz, showing the continuity of the combined trace and a good global accuracy of the relative frequency calibration. As the CO lines suffer the least Doppler broadening effect due to the largest molecular mass of CO, their frequencies are more precisely fitted with smaller deviations. Therefore to examine the precision of the relative frequency calibration, we fit the deviation distribution of CO lines with a Gaussian distribution and a 13.6 MHz standard deviation is shown. It should be emphasized that the 13.6 MHz standard deviation is not due to the method itself, but mainly due to the GHz Doppler broadening and the resulting higher statistical fitting error in determining the line-center frequencies. To eliminate the Doppler broadening and to demonstrate better precision, a Doppler-free spectroscopy could be built \cite{Nishiyama:13, Nishiyama:14}.



\begin{table}[b]
\centering
\caption{\bf Fitted dispersion parameters}
\begin{tabular}{ccc}
\hline
Parameter & TM$_{00}$ mode & TM$_{10}$ mode\\
\hline
$D_1/2\pi$ & 980.2 GHz  & 958.9 GHz\\
$D_2/2\pi$ & --26.08 MHz & --559.5 MHz\\
$D_3/2\pi$ & 5.62 MHz  & 46.3 MHz\\
$D_4/2\pi$ & --0.200 MHz & --2.03 MHz\\
\hline
\end{tabular}
\label{tab:DisParameter}
\end{table}

An important application of our method is to characterize the dispersion of microresonators. In this context, dispersion describes the variation of the FSR over frequency. It can be expressed in a Taylor series in analogue to the fiber case as

\begin{equation}
\begin{aligned}
\omega_{\mu}&=\omega_0+D_1\mu+\frac{1}{2}D_2\mu^2+\frac{1}{6}D_3\mu^3+\frac{1}{24}D_4\mu^4+...\\
&=\omega_0+D_1\mu+D_\text{int}(\mu)
\end{aligned}
\label{Eq:1}
\end{equation}

Here $\omega_{\mu}$ is the angular frequency of the $\mu$-th resonance relative to the reference resonance $\omega_0$. $D_1/2\pi$ corresponds to the FSR. A non-zero $D_2$ leads to a parabolic deviation from an equidistant $D_1$-spaced resonance grid. $D_\text{int}(\mu)$ is the integrated dispersion including $D_2$ and all higher order dispersion parameters, showing the total deviation.




We measured the dispersion of a Si$_3$N$_4$ microresonator fabricated using the photonic Damascene process \cite{Pfeiffer:16}. The microresonator has a radius of 23 $\mu m$ (corresponding $D_1/2\pi\approx 1$ THz) and is coupled by a single-mode bus waveguide, as shown in Fig. \ref{Fig:figure4}(a). Both the fundamental transverse magnetic mode (TM$_{00}$) and a higher order mode (TM$_{10}$) of the microresonator are excited by the TM waveguide mode. Referred to the simulations, they can be distinguished due to the different FSRs and linewidths as shown in Fig. \ref{Fig:figure4}(b) and \ref{Fig:figure4}(c). Each resonance is fitted with the model derived in Ref. \cite{Gorodetsky:00}, and the center frequency and the linewidth are extracted. Fig. \ref{Fig:figure4}(d) plots the $D_\text{int}/2\pi$ of the TM$_{00}$ mode, directly showing a visible $D_3$ contribution. Both a 3$^\text{rd}$ order and a 4$^\text{th}$ order weighted polynomial are used to fit the $D_\text{int}/2\pi$, while the reference resonance $\omega_0/2\pi$ is chosen at 193.12 THz (1553.5 nm). The fit is performed by weighting each resonance according to the inverse of its linewidth, as the center frequencies of broader resonances are less precisely fitted. As shown in Fig. \ref{Fig:figure4}(d), the 4$^\text{th}$ order weighted polynomial shows better fitting than the 3$^\text{rd}$ order one, indicating the necessity to consider the fourth order dispersion parameter $D_4$ in the fit. The dispersion parameters extracted from the 4$^\text{th}$ order polynomial fit for both TM$_{00}$ and TM$_{10}$ modes are shown in Table. \ref{tab:DisParameter}.

Fig. \ref{Fig:figure4}(e) plots the fitted linewidths of the TM$_{00}$ mode. The global trend showing decreasing linewidth with increasing frequency is due to the fact that the waveguide-resonator external coupling strength is wavelength-dependent \cite{Cai:00}. In addition, an avoided modal crossing \cite{Herr:14, Kordts:16} is identified around 213 THz where both modes have approximately the same local resonance frequency. At such a modal crossing point, the resonances deviate from the fitted dispersion curve and a local linewidth broadening is observed.

In conclusion, we have presented a novel way to extend frequency comb assisted diode laser spectroscopy with cascaded lasers, enabling a full measurement bandwidth of 37.4 THz (1355 -- 1630 nm) at MHz resolution with scanning speeds above 1 THz/s. We show potential applications for molecular spectroscopy and for the measurement of higher order dispersion in microresonators which we demonstrate here for the first time in a Si$_3$N$_4$ photonic chip-based microresonator. Furthermore the described cascaded laser spectroscopy can be extended with more lasers, enabling a further increase of the measurement bandwidth.


\section*{Funding Information}
We gratefully acknowledge funding via Defense Sciences Office (DSO), DARPA (W911NF-11-1-0202); European Space Agency (ESA) (ESTEC CN 4000105962/12/NL/PA); Swiss National Science Foundation (SNSF) (Schweizerischer Nationalfonds zur F\"orderung der Wissenschaftlichen Forschung (SNF)). M.G. acknowledges support from the EPFL fellowship programme co-funded by Marie Curie, FP7 Grant agreement no. 291771.

\section*{Acknowledgments}
The Si$_3$N$_4$ microresonator samples were fabricated in the EPFL center of MicroNanoTechnology (CMi).



\begin{thebibliography}{38}%
\makeatletter
\providecommand \@ifxundefined [1]{%
 \@ifx{#1\undefined}
}%
\providecommand \@ifnum [1]{%
 \ifnum #1\expandafter \@firstoftwo
 \else \expandafter \@secondoftwo
 \fi
}%
\providecommand \@ifx [1]{%
 \ifx #1\expandafter \@firstoftwo
 \else \expandafter \@secondoftwo
 \fi
}%
\providecommand \natexlab [1]{#1}%
\providecommand \enquote  [1]{``#1''}%
\providecommand \bibnamefont  [1]{#1}%
\providecommand \bibfnamefont [1]{#1}%
\providecommand \citenamefont [1]{#1}%
\providecommand \href@noop [0]{\@secondoftwo}%
\providecommand \href [0]{\begingroup \@sanitize@url \@href}%
\providecommand \@href[1]{\@@startlink{#1}\@@href}%
\providecommand \@@href[1]{\endgroup#1\@@endlink}%
\providecommand \@sanitize@url [0]{\catcode `\\12\catcode `\$12\catcode
  `\&12\catcode `\#12\catcode `\^12\catcode `\_12\catcode `\%12\relax}%
\providecommand \@@startlink[1]{}%
\providecommand \@@endlink[0]{}%
\providecommand \url  [0]{\begingroup\@sanitize@url \@url }%
\providecommand \@url [1]{\endgroup\@href {#1}{\urlprefix }}%
\providecommand \urlprefix  [0]{URL }%
\providecommand \Eprint [0]{\href }%
\providecommand \doibase [0]{http://dx.doi.org/}%
\providecommand \selectlanguage [0]{\@gobble}%
\providecommand \bibinfo  [0]{\@secondoftwo}%
\providecommand \bibfield  [0]{\@secondoftwo}%
\providecommand \translation [1]{[#1]}%
\providecommand \BibitemOpen [0]{}%
\providecommand \bibitemStop [0]{}%
\providecommand \bibitemNoStop [0]{.\EOS\space}%
\providecommand \EOS [0]{\spacefactor3000\relax}%
\providecommand \BibitemShut  [1]{\csname bibitem#1\endcsname}%
\let\auto@bib@innerbib\@empty
\bibitem [{\citenamefont {Udem}\ \emph {et~al.}(2002)\citenamefont {Udem},
  \citenamefont {Holzwarth},\ and\ \citenamefont {H\"ansch}}]{Udem:02}%
  \BibitemOpen
  \bibfield  {author} {\bibinfo {author} {\bibfnamefont {T.}~\bibnamefont
  {Udem}}, \bibinfo {author} {\bibfnamefont {R.}~\bibnamefont {Holzwarth}}, \
  and\ \bibinfo {author} {\bibfnamefont {T.~W.}\ \bibnamefont {H\"ansch}},\
  }\href {http://dx.doi.org/10.1038/416233a} {\bibfield  {journal} {\bibinfo
  {journal} {Nature}\ }\textbf {\bibinfo {volume} {416}},\ \bibinfo {pages}
  {233} (\bibinfo {year} {2002})}\BibitemShut {NoStop}%
\bibitem [{\citenamefont {Cundiff}\ and\ \citenamefont
  {Ye}(2003)}]{Cundiff:03}%
  \BibitemOpen
  \bibfield  {author} {\bibinfo {author} {\bibfnamefont {S.~T.}\ \bibnamefont
  {Cundiff}}\ and\ \bibinfo {author} {\bibfnamefont {J.}~\bibnamefont {Ye}},\
  }\href {\doibase 10.1103/RevModPhys.75.325} {\bibfield  {journal} {\bibinfo
  {journal} {Rev. Mod. Phys.}\ }\textbf {\bibinfo {volume} {75}},\ \bibinfo
  {pages} {325} (\bibinfo {year} {2003})}\BibitemShut {NoStop}%

\bibitem [{\citenamefont {Foltynowicz}\ \emph {et~al.}(2011)\citenamefont
  {Foltynowicz}, \citenamefont {Maslowski}, \citenamefont {Ban}, \citenamefont
  {Adler}, \citenamefont {Cossel}, \citenamefont {Briles},\ and\ \citenamefont
  {Ye}}]{Foltynowicz:11}%
  \BibitemOpen
  \bibfield  {author} {\bibinfo {author} {\bibfnamefont {A.}~\bibnamefont
  {Foltynowicz}}, \bibinfo {author} {\bibfnamefont {P.}~\bibnamefont
  {Maslowski}}, \bibinfo {author} {\bibfnamefont {T.}~\bibnamefont {Ban}},
  \bibinfo {author} {\bibfnamefont {F.}~\bibnamefont {Adler}}, \bibinfo
  {author} {\bibfnamefont {K.}~\bibnamefont {Cossel}}, \bibinfo {author}
  {\bibfnamefont {T.}~\bibnamefont {Briles}}, \ and\ \bibinfo {author}
  {\bibfnamefont {J.}~\bibnamefont {Ye}},\ }\href {http://dx.doi.org/10.1039/C1FD00005E} {\bibfield  {journal}
  {\bibinfo  {journal} {Faraday Discussions}\ }\textbf {\bibinfo {volume}
  {150}},\ \bibinfo {pages} {23} (\bibinfo {year} {2011})}\BibitemShut
  {NoStop}%
\bibitem [{\citenamefont {Mandon}\ \emph {et~al.}(2009)\citenamefont {Mandon},
  \citenamefont {Guelachvili},\ and\ \citenamefont {Picque}}]{Mandon:09}%
  \BibitemOpen
  \bibfield  {author} {\bibinfo {author} {\bibfnamefont {J.}~\bibnamefont
  {Mandon}}, \bibinfo {author} {\bibfnamefont {G.}~\bibnamefont {Guelachvili}},
  \ and\ \bibinfo {author} {\bibfnamefont {N.}~\bibnamefont {Picque}},\ }\href
  {http://dx.doi.org/10.1038/nphoton.2008.293} {\bibfield  {journal} {\bibinfo
  {journal} {Nat Photon}\ }\textbf {\bibinfo {volume} {3}},\ \bibinfo {pages}
  {99} (\bibinfo {year} {2009})}\BibitemShut {NoStop}%
\bibitem [{\citenamefont {Bernhardt}\ \emph {et~al.}(2010)\citenamefont
  {Bernhardt}, \citenamefont {Ozawa}, \citenamefont {Jacquet}, \citenamefont
  {Jacquey}, \citenamefont {Kobayashi}, \citenamefont {Udem}, \citenamefont
  {Holzwarth}, \citenamefont {Guelachvili}, \citenamefont {H\"ansch},\ and\
  \citenamefont {Picque}}]{Bernhardt:10}%
  \BibitemOpen
  \bibfield  {author} {\bibinfo {author} {\bibfnamefont {B.}~\bibnamefont
  {Bernhardt}}, \bibinfo {author} {\bibfnamefont {A.}~\bibnamefont {Ozawa}},
  \bibinfo {author} {\bibfnamefont {P.}~\bibnamefont {Jacquet}}, \bibinfo
  {author} {\bibfnamefont {M.}~\bibnamefont {Jacquey}}, \bibinfo {author}
  {\bibfnamefont {Y.}~\bibnamefont {Kobayashi}}, \bibinfo {author}
  {\bibfnamefont {T.}~\bibnamefont {Udem}}, \bibinfo {author} {\bibfnamefont
  {R.}~\bibnamefont {Holzwarth}}, \bibinfo {author} {\bibfnamefont
  {G.}~\bibnamefont {Guelachvili}}, \bibinfo {author} {\bibfnamefont {T.~W.}\
  \bibnamefont {H\"ansch}}, \ and\ \bibinfo {author} {\bibfnamefont
  {N.}~\bibnamefont {Picque}},\ }\href
  {http://dx.doi.org/10.1038/nphoton.2009.217} {\bibfield  {journal} {\bibinfo
  {journal} {Nat Photon}\ }\textbf {\bibinfo {volume} {4}},\ \bibinfo {pages}
  {55} (\bibinfo {year} {2010})}\BibitemShut {NoStop}%
\bibitem [{\citenamefont {Del'Haye}\ \emph {et~al.}(2009)\citenamefont
  {Del'Haye}, \citenamefont {Arcizet}, \citenamefont {Gorodetsky},
  \citenamefont {Holzwarth},\ and\ \citenamefont {Kippenberg}}]{DelHaye:09}%
  \BibitemOpen
  \bibfield  {author} {\bibinfo {author} {\bibfnamefont {P.}~\bibnamefont
  {Del'Haye}}, \bibinfo {author} {\bibfnamefont {O.}~\bibnamefont {Arcizet}},
  \bibinfo {author} {\bibfnamefont {M.~L.}\ \bibnamefont {Gorodetsky}},
  \bibinfo {author} {\bibfnamefont {R.}~\bibnamefont {Holzwarth}}, \ and\
  \bibinfo {author} {\bibfnamefont {T.~J.}\ \bibnamefont {Kippenberg}},\ }\href
  {http://dx.doi.org/10.1038/nphoton.2009.138} {\bibfield  {journal} {\bibinfo
  {journal} {Nat Photon}\ }\textbf {\bibinfo {volume} {3}},\ \bibinfo {pages}
  {529} (\bibinfo {year} {2009})}\BibitemShut {NoStop}%
\bibitem [{\citenamefont {Baumann}\ \emph {et~al.}(2013)\citenamefont
  {Baumann}, \citenamefont {Giorgetta}, \citenamefont {Coddington},
  \citenamefont {Sinclair}, \citenamefont {Knabe}, \citenamefont {Swann},\ and\
  \citenamefont {Newbury}}]{Baumann:13}%
  \BibitemOpen
  \bibfield  {author} {\bibinfo {author} {\bibfnamefont {E.}~\bibnamefont
  {Baumann}}, \bibinfo {author} {\bibfnamefont {F.~R.}\ \bibnamefont
  {Giorgetta}}, \bibinfo {author} {\bibfnamefont {I.}~\bibnamefont
  {Coddington}}, \bibinfo {author} {\bibfnamefont {L.~C.}\ \bibnamefont
  {Sinclair}}, \bibinfo {author} {\bibfnamefont {K.}~\bibnamefont {Knabe}},
  \bibinfo {author} {\bibfnamefont {W.~C.}\ \bibnamefont {Swann}}, \ and\
  \bibinfo {author} {\bibfnamefont {N.~R.}\ \bibnamefont {Newbury}},\ }\href
  {\doibase 10.1364/OL.38.002026} {\bibfield  {journal} {\bibinfo  {journal}
  {Opt. Lett.}\ }\textbf {\bibinfo {volume} {38}},\ \bibinfo {pages} {2026}
  (\bibinfo {year} {2013})}\BibitemShut {NoStop}%
\bibitem [{\citenamefont {Baumann}\ \emph {et~al.}(2014)\citenamefont
  {Baumann}, \citenamefont {Giorgetta}, \citenamefont {Desch\^{e}nes},
  \citenamefont {Swann}, \citenamefont {Coddington},\ and\ \citenamefont
  {Newbury}}]{Baumann:14}%
  \BibitemOpen
  \bibfield  {author} {\bibinfo {author} {\bibfnamefont {E.}~\bibnamefont
  {Baumann}}, \bibinfo {author} {\bibfnamefont {F.~R.}\ \bibnamefont
  {Giorgetta}}, \bibinfo {author} {\bibfnamefont {J.-D.}\ \bibnamefont
  {Desch\^{e}nes}}, \bibinfo {author} {\bibfnamefont {W.~C.}\ \bibnamefont
  {Swann}}, \bibinfo {author} {\bibfnamefont {I.}~\bibnamefont {Coddington}}, \
  and\ \bibinfo {author} {\bibfnamefont {N.~R.}\ \bibnamefont {Newbury}},\
  }\href {\doibase 10.1364/OE.22.024914} {\bibfield  {journal} {\bibinfo
  {journal} {Opt. Express}\ }\textbf {\bibinfo {volume} {22}},\ \bibinfo
  {pages} {24914} (\bibinfo {year} {2014})}\BibitemShut {NoStop}%
\bibitem [{\citenamefont {Giorgetta}\ \emph {et~al.}(2010)\citenamefont
  {Giorgetta}, \citenamefont {Coddington}, \citenamefont {Baumann},
  \citenamefont {Swann},\ and\ \citenamefont {Newbury}}]{Giorgetta:10}%
  \BibitemOpen
  \bibfield  {author} {\bibinfo {author} {\bibfnamefont {F.~R.}\ \bibnamefont
  {Giorgetta}}, \bibinfo {author} {\bibfnamefont {I.}~\bibnamefont
  {Coddington}}, \bibinfo {author} {\bibfnamefont {E.}~\bibnamefont {Baumann}},
  \bibinfo {author} {\bibfnamefont {W.~C.}\ \bibnamefont {Swann}}, \ and\
  \bibinfo {author} {\bibfnamefont {N.~R.}\ \bibnamefont {Newbury}},\ }\href
  {http://dx.doi.org/10.1038/nphoton.2010.228} {\bibfield  {journal} {\bibinfo
  {journal} {Nat Photon}\ }\textbf {\bibinfo {volume} {4}},\ \bibinfo {pages}
  {853} (\bibinfo {year} {2010})}\BibitemShut {NoStop}%
\bibitem [{\citenamefont {Urabe}\ and\ \citenamefont {Sakai}(2012)}]{Urabe:12}%
  \BibitemOpen
  \bibfield  {author} {\bibinfo {author} {\bibfnamefont {K.}~\bibnamefont
  {Urabe}}\ and\ \bibinfo {author} {\bibfnamefont {O.}~\bibnamefont {Sakai}},\
  }\href {\doibase 10.1063/1.4742136} {\bibfield  {journal} {\bibinfo
  {journal} {Appl. Phys. Lett.}\ }\textbf {\bibinfo {volume} {101}},\
  \bibinfo {eid} {051105} (\bibinfo {year} {2012})}\BibitemShut {NoStop}%
\bibitem [{\citenamefont {Nishiyama}\ \emph {et~al.}(2013)\citenamefont
  {Nishiyama}, \citenamefont {Ishikawa},\ and\ \citenamefont
  {Misono}}]{Nishiyama:13}%
  \BibitemOpen
  \bibfield  {author} {\bibinfo {author} {\bibfnamefont {A.}~\bibnamefont
  {Nishiyama}}, \bibinfo {author} {\bibfnamefont {D.}~\bibnamefont {Ishikawa}},
  \ and\ \bibinfo {author} {\bibfnamefont {M.}~\bibnamefont {Misono}},\ }\href
  {\doibase 10.1364/JOSAB.30.002107} {\bibfield  {journal} {\bibinfo  {journal}
  {J. Opt. Soc. Am. B}\ }\textbf {\bibinfo {volume} {30}},\ \bibinfo {pages}
  {2107} (\bibinfo {year} {2013})}\BibitemShut {NoStop}%
\bibitem [{\citenamefont {Nishiyama}\ \emph {et~al.}(2014)\citenamefont
  {Nishiyama}, \citenamefont {Matsuba},\ and\ \citenamefont
  {Misono}}]{Nishiyama:14}%
  \BibitemOpen
  \bibfield  {author} {\bibinfo {author} {\bibfnamefont {A.}~\bibnamefont
  {Nishiyama}}, \bibinfo {author} {\bibfnamefont {A.}~\bibnamefont {Matsuba}},
  \ and\ \bibinfo {author} {\bibfnamefont {M.}~\bibnamefont {Misono}},\ }\href
  {\doibase 10.1364/OL.39.004923} {\bibfield  {journal} {\bibinfo  {journal}
  {Opt. Lett.}\ }\textbf {\bibinfo {volume} {39}},\ \bibinfo {pages} {4923}
  (\bibinfo {year} {2014})}\BibitemShut {NoStop}%
\bibitem [{\citenamefont {Riemensberger}\ \emph {et~al.}(2012)\citenamefont
  {Riemensberger}, \citenamefont {Hartinger}, \citenamefont {Herr},
  \citenamefont {Brasch}, \citenamefont {Holzwarth},\ and\ \citenamefont
  {Kippenberg}}]{Riemensberger:12}%
  \BibitemOpen
  \bibfield  {author} {\bibinfo {author} {\bibfnamefont {J.}~\bibnamefont
  {Riemensberger}}, \bibinfo {author} {\bibfnamefont {K.}~\bibnamefont
  {Hartinger}}, \bibinfo {author} {\bibfnamefont {T.}~\bibnamefont {Herr}},
  \bibinfo {author} {\bibfnamefont {V.}~\bibnamefont {Brasch}}, \bibinfo
  {author} {\bibfnamefont {R.}~\bibnamefont {Holzwarth}}, \ and\ \bibinfo
  {author} {\bibfnamefont {T.~J.}\ \bibnamefont {Kippenberg}},\ }\href
  {\doibase 10.1364/OE.20.027661} {\bibfield  {journal} {\bibinfo  {journal}
  {Opt. Express}\ }\textbf {\bibinfo {volume} {20}},\ \bibinfo {pages} {27661}
  (\bibinfo {year} {2012})}\BibitemShut {NoStop}%
\bibitem [{\citenamefont {Herr}\ \emph
  {et~al.}(2014{\natexlab{a}})\citenamefont {Herr}, \citenamefont {Brasch},
  \citenamefont {Jost}, \citenamefont {Mirgorodskiy}, \citenamefont {Lihachev},
  \citenamefont {Gorodetsky},\ and\ \citenamefont {Kippenberg}}]{Herr:14}%
  \BibitemOpen
  \bibfield  {author} {\bibinfo {author} {\bibfnamefont {T.}~\bibnamefont
  {Herr}}, \bibinfo {author} {\bibfnamefont {V.}~\bibnamefont {Brasch}},
  \bibinfo {author} {\bibfnamefont {J.~D.}\ \bibnamefont {Jost}}, \bibinfo
  {author} {\bibfnamefont {I.}~\bibnamefont {Mirgorodskiy}}, \bibinfo {author}
  {\bibfnamefont {G.}~\bibnamefont {Lihachev}}, \bibinfo {author}
  {\bibfnamefont {M.~L.}\ \bibnamefont {Gorodetsky}}, \ and\ \bibinfo {author}
  {\bibfnamefont {T.~J.}\ \bibnamefont {Kippenberg}},\ }\href {\doibase
  10.1103/PhysRevLett.113.123901} {\bibfield  {journal} {\bibinfo  {journal}
  {Phys. Rev. Lett.}\ }\textbf {\bibinfo {volume} {113}},\ \bibinfo {pages}
  {123901} (\bibinfo {year} {2014}{\natexlab{a}})}\BibitemShut {NoStop}%
\bibitem [{\citenamefont {Kordts}\ \emph {et~al.}(2016)\citenamefont {Kordts},
  \citenamefont {Pfeiffer}, \citenamefont {Guo}, \citenamefont {Brasch},\ and\
  \citenamefont {Kippenberg}}]{Kordts:16}%
  \BibitemOpen
  \bibfield  {author} {\bibinfo {author} {\bibfnamefont {A.}~\bibnamefont
  {Kordts}}, \bibinfo {author} {\bibfnamefont {M.~H.~P.}\ \bibnamefont
  {Pfeiffer}}, \bibinfo {author} {\bibfnamefont {H.}~\bibnamefont {Guo}},
  \bibinfo {author} {\bibfnamefont {V.}~\bibnamefont {Brasch}}, \ and\ \bibinfo
  {author} {\bibfnamefont {T.~J.}\ \bibnamefont {Kippenberg}},\ }\href
  {\doibase 10.1364/OL.41.000452} {\bibfield  {journal} {\bibinfo  {journal}
  {Opt. Lett.}\ }\textbf {\bibinfo {volume} {41}},\ \bibinfo {pages} {452}
  (\bibinfo {year} {2016})}\BibitemShut {NoStop}%
\bibitem [{\citenamefont {Del'Haye}\ \emph {et~al.}(2015)\citenamefont
  {Del'Haye}, \citenamefont {Coillet}, \citenamefont {Loh}, \citenamefont
  {Beha}, \citenamefont {Papp},\ and\ \citenamefont {Diddams}}]{DelHaye:15}%
  \BibitemOpen
  \bibfield  {author} {\bibinfo {author} {\bibfnamefont {P.}~\bibnamefont
  {Del'Haye}}, \bibinfo {author} {\bibfnamefont {A.}~\bibnamefont {Coillet}},
  \bibinfo {author} {\bibfnamefont {W.}~\bibnamefont {Loh}}, \bibinfo {author}
  {\bibfnamefont {K.}~\bibnamefont {Beha}}, \bibinfo {author} {\bibfnamefont
  {S.~B.}\ \bibnamefont {Papp}}, \ and\ \bibinfo {author} {\bibfnamefont
  {S.~A.}\ \bibnamefont {Diddams}},\ }\href
  {http://dx.doi.org/10.1038/ncomms6668} {\bibfield  {journal} {\bibinfo
  {journal} {Nat Commun}\ }\textbf {\bibinfo {volume} {6}} (\bibinfo {year}
  {2015})}\BibitemShut {NoStop}%
\bibitem [{\citenamefont {Pfeiffer}\ \emph {et~al.}(2016)\citenamefont
  {Pfeiffer}, \citenamefont {Kordts}, \citenamefont {Brasch}, \citenamefont
  {Zervas}, \citenamefont {Geiselmann}, \citenamefont {Jost},\ and\
  \citenamefont {Kippenberg}}]{Pfeiffer:16}%
  \BibitemOpen
  \bibfield  {author} {\bibinfo {author} {\bibfnamefont {M.~H.~P.}\
  \bibnamefont {Pfeiffer}}, \bibinfo {author} {\bibfnamefont {A.}~\bibnamefont
  {Kordts}}, \bibinfo {author} {\bibfnamefont {V.}~\bibnamefont {Brasch}},
  \bibinfo {author} {\bibfnamefont {M.}~\bibnamefont {Zervas}}, \bibinfo
  {author} {\bibfnamefont {M.}~\bibnamefont {Geiselmann}}, \bibinfo {author}
  {\bibfnamefont {J.~D.}\ \bibnamefont {Jost}}, \ and\ \bibinfo {author}
  {\bibfnamefont {T.~J.}\ \bibnamefont {Kippenberg}},\ }\href {\doibase
  10.1364/OPTICA.3.000020} {\bibfield  {journal} {\bibinfo  {journal} {Optica}\
  }\textbf {\bibinfo {volume} {3}},\ \bibinfo {pages} {20} (\bibinfo {year}
  {2016})}\BibitemShut {NoStop}%
\bibitem [{\citenamefont {Thorpe}\ \emph {et~al.}(2005)\citenamefont {Thorpe},
  \citenamefont {Jones}, \citenamefont {Moll}, \citenamefont {Ye},\ and\
  \citenamefont {Lalezari}}]{Thorpe:05}%
  \BibitemOpen
  \bibfield  {author} {\bibinfo {author} {\bibfnamefont {M.~J.}\ \bibnamefont
  {Thorpe}}, \bibinfo {author} {\bibfnamefont {R.~J.}\ \bibnamefont {Jones}},
  \bibinfo {author} {\bibfnamefont {K.~D.}\ \bibnamefont {Moll}}, \bibinfo
  {author} {\bibfnamefont {J.}~\bibnamefont {Ye}}, \ and\ \bibinfo {author}
  {\bibfnamefont {R.}~\bibnamefont {Lalezari}},\ }\href {\doibase
  10.1364/OPEX.13.000882} {\bibfield  {journal} {\bibinfo  {journal} {Opt.
  Express}\ }\textbf {\bibinfo {volume} {13}},\ \bibinfo {pages} {882}
  (\bibinfo {year} {2005})}\BibitemShut {NoStop}%
\bibitem [{\citenamefont {Schliesser}\ \emph {et~al.}(2006)\citenamefont
  {Schliesser}, \citenamefont {Gohle}, \citenamefont {Udem},\ and\
  \citenamefont {H\"{a}nsch}}]{Schliesser:06}%
  \BibitemOpen
  \bibfield  {author} {\bibinfo {author} {\bibfnamefont {A.}~\bibnamefont
  {Schliesser}}, \bibinfo {author} {\bibfnamefont {C.}~\bibnamefont {Gohle}},
  \bibinfo {author} {\bibfnamefont {T.}~\bibnamefont {Udem}}, \ and\ \bibinfo
  {author} {\bibfnamefont {T.~W.}\ \bibnamefont {H\"{a}nsch}},\ }\href
  {\doibase 10.1364/OE.14.005975} {\bibfield  {journal} {\bibinfo  {journal}
  {Opt. Express}\ }\textbf {\bibinfo {volume} {14}},\ \bibinfo {pages} {5975}
  (\bibinfo {year} {2006})}\BibitemShut {NoStop}%
\bibitem [{\citenamefont {Savchenkov}\ \emph {et~al.}(2008)\citenamefont
  {Savchenkov}, \citenamefont {Rubiola}, \citenamefont {Matsko}, \citenamefont
  {Ilchenko},\ and\ \citenamefont {Maleki}}]{Savchenkov:08}%
  \BibitemOpen
  \bibfield  {author} {\bibinfo {author} {\bibfnamefont {A.~A.}\ \bibnamefont
  {Savchenkov}}, \bibinfo {author} {\bibfnamefont {E.}~\bibnamefont {Rubiola}},
  \bibinfo {author} {\bibfnamefont {A.~B.}\ \bibnamefont {Matsko}}, \bibinfo
  {author} {\bibfnamefont {V.~S.}\ \bibnamefont {Ilchenko}}, \ and\ \bibinfo
  {author} {\bibfnamefont {L.}~\bibnamefont {Maleki}},\ }\href {\doibase
  10.1364/OE.16.004130} {\bibfield  {journal} {\bibinfo  {journal} {Opt.
  Express}\ }\textbf {\bibinfo {volume} {16}},\ \bibinfo {pages} {4130}
  (\bibinfo {year} {2008})}\BibitemShut {NoStop}%
\bibitem [{\citenamefont {Li}\ \emph {et~al.}(2012)\citenamefont {Li},
  \citenamefont {Lee}, \citenamefont {Yang},\ and\ \citenamefont
  {Vahala}}]{Li:12}%
  \BibitemOpen
  \bibfield  {author} {\bibinfo {author} {\bibfnamefont {J.}~\bibnamefont
  {Li}}, \bibinfo {author} {\bibfnamefont {H.}~\bibnamefont {Lee}}, \bibinfo
  {author} {\bibfnamefont {K.~Y.}\ \bibnamefont {Yang}}, \ and\ \bibinfo
  {author} {\bibfnamefont {K.~J.}\ \bibnamefont {Vahala}},\ }\href {\doibase
  10.1364/OE.20.026337} {\bibfield  {journal} {\bibinfo  {journal} {Opt.
  Express}\ }\textbf {\bibinfo {volume} {20}},\ \bibinfo {pages} {26337}
  (\bibinfo {year} {2012})}\BibitemShut {NoStop}%
\bibitem [{\citenamefont {Del'Haye}\ \emph {et~al.}(2007)\citenamefont
  {Del'Haye}, \citenamefont {Schliesser}, \citenamefont {Arcizet},
  \citenamefont {Wilken}, \citenamefont {Holzwarth},\ and\ \citenamefont
  {Kippenberg}}]{DelHaye:07}%
  \BibitemOpen
  \bibfield  {author} {\bibinfo {author} {\bibfnamefont {P.}~\bibnamefont
  {Del'Haye}}, \bibinfo {author} {\bibfnamefont {A.}~\bibnamefont
  {Schliesser}}, \bibinfo {author} {\bibfnamefont {O.}~\bibnamefont {Arcizet}},
  \bibinfo {author} {\bibfnamefont {T.}~\bibnamefont {Wilken}}, \bibinfo
  {author} {\bibfnamefont {R.}~\bibnamefont {Holzwarth}}, \ and\ \bibinfo
  {author} {\bibfnamefont {T.~J.}\ \bibnamefont {Kippenberg}},\ }\href
  {http://dx.doi.org/10.1038/nature06401} {\bibfield  {journal} {\bibinfo
  {journal} {Nature}\ }\textbf {\bibinfo {volume} {450}},\ \bibinfo {pages}
  {1214} (\bibinfo {year} {2007})}\BibitemShut {NoStop}%
\bibitem [{\citenamefont {Kippenberg}\ \emph {et~al.}(2011)\citenamefont
  {Kippenberg}, \citenamefont {Holzwarth},\ and\ \citenamefont
  {Diddams}}]{Kippenberg:11}%
  \BibitemOpen
  \bibfield  {author} {\bibinfo {author} {\bibfnamefont {T.~J.}\ \bibnamefont
  {Kippenberg}}, \bibinfo {author} {\bibfnamefont {R.}~\bibnamefont
  {Holzwarth}}, \ and\ \bibinfo {author} {\bibfnamefont {S.~A.}\ \bibnamefont
  {Diddams}},\ }\href {\doibase 10.1126/science.1193968} {\bibfield  {journal}
  {\bibinfo  {journal} {Science}\ }\textbf {\bibinfo {volume} {332}},\ \bibinfo
  {pages} {555} (\bibinfo {year} {2011})}\BibitemShut {NoStop}%
\bibitem [{\citenamefont {Herr}\ \emph
  {et~al.}(2014{\natexlab{b}})\citenamefont {Herr}, \citenamefont {Brasch},
  \citenamefont {Jost}, \citenamefont {Wang}, \citenamefont {Kondratiev},
  \citenamefont {Gorodetsky},\ and\ \citenamefont {Kippenberg}}]{HerrNP:14}%
  \BibitemOpen
  \bibfield  {author} {\bibinfo {author} {\bibfnamefont {T.}~\bibnamefont
  {Herr}}, \bibinfo {author} {\bibfnamefont {V.}~\bibnamefont {Brasch}},
  \bibinfo {author} {\bibfnamefont {J.~D.}\ \bibnamefont {Jost}}, \bibinfo
  {author} {\bibfnamefont {C.~Y.}\ \bibnamefont {Wang}}, \bibinfo {author}
  {\bibfnamefont {N.~M.}\ \bibnamefont {Kondratiev}}, \bibinfo {author}
  {\bibfnamefont {M.~L.}\ \bibnamefont {Gorodetsky}}, \ and\ \bibinfo {author}
  {\bibfnamefont {T.~J.}\ \bibnamefont {Kippenberg}},\ }\href
  {http://dx.doi.org/10.1038/nphoton.2013.343} {\bibfield  {journal} {\bibinfo
  {journal} {Nat Photon}\ }\textbf {\bibinfo {volume} {8}},\ \bibinfo {pages}
  {145} (\bibinfo {year} {2014}{\natexlab{b}})}\BibitemShut {NoStop}%
\bibitem [{\citenamefont {Yi}\ \emph {et~al.}(2015)\citenamefont {Yi},
  \citenamefont {Yang}, \citenamefont {Yang}, \citenamefont {Suh},\ and\
  \citenamefont {Vahala}}]{Yi:15}%
  \BibitemOpen
  \bibfield  {author} {\bibinfo {author} {\bibfnamefont {X.}~\bibnamefont
  {Yi}}, \bibinfo {author} {\bibfnamefont {Q.-F.}\ \bibnamefont {Yang}},
  \bibinfo {author} {\bibfnamefont {K.~Y.}\ \bibnamefont {Yang}}, \bibinfo
  {author} {\bibfnamefont {M.-G.}\ \bibnamefont {Suh}}, \ and\ \bibinfo
  {author} {\bibfnamefont {K.}~\bibnamefont {Vahala}},\ }\href {\doibase
  10.1364/OPTICA.2.001078} {\bibfield  {journal} {\bibinfo  {journal} {Optica}\
  }\textbf {\bibinfo {volume} {2}},\ \bibinfo {pages} {1078} (\bibinfo {year}
  {2015})}\BibitemShut {NoStop}%
\bibitem [{\citenamefont {Joshi}\ \emph {et~al.}(2016)\citenamefont {Joshi},
  \citenamefont {Jang}, \citenamefont {Luke}, \citenamefont {Ji}, \citenamefont
  {Miller}, \citenamefont {Klenner}, \citenamefont {Okawachi}, \citenamefont
  {Lipson},\ and\ \citenamefont {Gaeta}}]{Joshi:16}%
  \BibitemOpen
  \bibfield  {author} {\bibinfo {author} {\bibfnamefont {C.}~\bibnamefont
  {Joshi}}, \bibinfo {author} {\bibfnamefont {J.~K.}\ \bibnamefont {Jang}},
  \bibinfo {author} {\bibfnamefont {K.}~\bibnamefont {Luke}}, \bibinfo {author}
  {\bibfnamefont {X.}~\bibnamefont {Ji}}, \bibinfo {author} {\bibfnamefont
  {S.~A.}\ \bibnamefont {Miller}}, \bibinfo {author} {\bibfnamefont
  {A.}~\bibnamefont {Klenner}}, \bibinfo {author} {\bibfnamefont
  {Y.}~\bibnamefont {Okawachi}}, \bibinfo {author} {\bibfnamefont
  {M.}~\bibnamefont {Lipson}}, \ and\ \bibinfo {author} {\bibfnamefont {A.~L.}\
  \bibnamefont {Gaeta}},\ }\href@noop {} {\bibfield  {journal} {\bibinfo
  {journal} {arXiv:1603.08017}\ } (\bibinfo {year} {2016})}\BibitemShut
  {NoStop}%
\bibitem [{\citenamefont {Brasch}\ \emph {et~al.}(2015)\citenamefont {Brasch},
  \citenamefont {Geiselmann}, \citenamefont {Herr}, \citenamefont {Lihachev},
  \citenamefont {Pfeiffer}, \citenamefont {Gorodetsky},\ and\ \citenamefont
  {Kippenberg}}]{Brasch:15}%
  \BibitemOpen
  \bibfield  {author} {\bibinfo {author} {\bibfnamefont {V.}~\bibnamefont
  {Brasch}}, \bibinfo {author} {\bibfnamefont {M.}~\bibnamefont {Geiselmann}},
  \bibinfo {author} {\bibfnamefont {T.}~\bibnamefont {Herr}}, \bibinfo {author}
  {\bibfnamefont {G.}~\bibnamefont {Lihachev}}, \bibinfo {author}
  {\bibfnamefont {M.~H.~P.}\ \bibnamefont {Pfeiffer}}, \bibinfo {author}
  {\bibfnamefont {M.~L.}\ \bibnamefont {Gorodetsky}}, \ and\ \bibinfo {author}
  {\bibfnamefont {T.~J.}\ \bibnamefont {Kippenberg}},\ }\href {\doibase
  http://dx.doi.org/10.1126/science.aad4811} {\bibfield  {journal} {\bibinfo
  {journal} {Science}\ } (\bibinfo {year} {2015})}\BibitemShut {NoStop}%
\bibitem [{\citenamefont {Mili\'{a}n}\ and\ \citenamefont
  {Skryabin}(2014)}]{Milian:14}%
  \BibitemOpen
  \bibfield  {author} {\bibinfo {author} {\bibfnamefont {C.}~\bibnamefont
  {Mili\'{a}n}}\ and\ \bibinfo {author} {\bibfnamefont {D.}~\bibnamefont
  {Skryabin}},\ }\href {\doibase 10.1364/OE.22.003732} {\bibfield  {journal}
  {\bibinfo  {journal} {Opt. Express}\ }\textbf {\bibinfo {volume} {22}},\
  \bibinfo {pages} {3732} (\bibinfo {year} {2014})}\BibitemShut {NoStop}%
\bibitem [{\citenamefont {Jang}\ \emph {et~al.}(2014)\citenamefont {Jang},
  \citenamefont {Erkintalo}, \citenamefont {Murdoch},\ and\ \citenamefont
  {Coen}}]{Jang:14}%
  \BibitemOpen
  \bibfield  {author} {\bibinfo {author} {\bibfnamefont {J.~K.}\ \bibnamefont
  {Jang}}, \bibinfo {author} {\bibfnamefont {M.}~\bibnamefont {Erkintalo}},
  \bibinfo {author} {\bibfnamefont {S.~G.}\ \bibnamefont {Murdoch}}, \ and\
  \bibinfo {author} {\bibfnamefont {S.}~\bibnamefont {Coen}},\ }\href {\doibase
  10.1364/OL.39.005503} {\bibfield  {journal} {\bibinfo  {journal} {Opt.
  Lett.}\ }\textbf {\bibinfo {volume} {39}},\ \bibinfo {pages} {5503} (\bibinfo
  {year} {2014})}\BibitemShut {NoStop}%
\bibitem [{\citenamefont {Karpov}\ \emph {et~al.}(2016)\citenamefont {Karpov},
  \citenamefont {Guo}, \citenamefont {Kordts}, \citenamefont {Brasch},
  \citenamefont {Pfeiffer}, \citenamefont {Zervas}, \citenamefont
  {Geiselmann},\ and\ \citenamefont {Kippenberg}}]{Karpov:16}%
  \BibitemOpen
  \bibfield  {author} {\bibinfo {author} {\bibfnamefont {M.}~\bibnamefont
  {Karpov}}, \bibinfo {author} {\bibfnamefont {H.}~\bibnamefont {Guo}},
  \bibinfo {author} {\bibfnamefont {A.}~\bibnamefont {Kordts}}, \bibinfo
  {author} {\bibfnamefont {V.}~\bibnamefont {Brasch}}, \bibinfo {author}
  {\bibfnamefont {M.~H.~P.}\ \bibnamefont {Pfeiffer}}, \bibinfo {author}
  {\bibfnamefont {M.}~\bibnamefont {Zervas}}, \bibinfo {author} {\bibfnamefont
  {M.}~\bibnamefont {Geiselmann}}, \ and\ \bibinfo {author} {\bibfnamefont
  {T.~J.}\ \bibnamefont {Kippenberg}},\ }\href {\doibase
  10.1103/PhysRevLett.116.103902} {\bibfield  {journal} {\bibinfo  {journal}
  {Phys. Rev. Lett.}\ }\textbf {\bibinfo {volume} {116}},\ \bibinfo {pages}
  {103902} (\bibinfo {year} {2016})}\BibitemShut {NoStop}%
\bibitem [{\citenamefont {Yang}\ \emph {et~al.}(2016)\citenamefont {Yang},
  \citenamefont {Beha}, \citenamefont {Cole}, \citenamefont {Yi}, \citenamefont
  {Del'Haye}, \citenamefont {Lee}, \citenamefont {Li}, \citenamefont {Oh},
  \citenamefont {Diddams}, \citenamefont {Papp},\ and\ \citenamefont
  {Vahala}}]{Yang:16}%
  \BibitemOpen
  \bibfield  {author} {\bibinfo {author} {\bibfnamefont {K.~Y.}\ \bibnamefont
  {Yang}}, \bibinfo {author} {\bibfnamefont {K.}~\bibnamefont {Beha}}, \bibinfo
  {author} {\bibfnamefont {D.~C.}\ \bibnamefont {Cole}}, \bibinfo {author}
  {\bibfnamefont {X.}~\bibnamefont {Yi}}, \bibinfo {author} {\bibfnamefont
  {P.}~\bibnamefont {Del'Haye}}, \bibinfo {author} {\bibfnamefont
  {H.}~\bibnamefont {Lee}}, \bibinfo {author} {\bibfnamefont {J.}~\bibnamefont
  {Li}}, \bibinfo {author} {\bibfnamefont {D.~Y.}\ \bibnamefont {Oh}}, \bibinfo
  {author} {\bibfnamefont {S.~A.}\ \bibnamefont {Diddams}}, \bibinfo {author}
  {\bibfnamefont {S.~B.}\ \bibnamefont {Papp}}, \ and\ \bibinfo {author}
  {\bibfnamefont {K.~J.}\ \bibnamefont {Vahala}},\ }\href
  {http://dx.doi.org/10.1038/nphoton.2016.36} {\bibfield  {journal} {\bibinfo
  {journal} {Nat Photon}\ }\textbf {\bibinfo {volume} {advance online
  publication}},\  (\bibinfo {year} {2016})}\BibitemShut {NoStop}%
\bibitem [{\citenamefont {Jiang}\ \emph {et~al.}(2014)\citenamefont {Jiang},
  \citenamefont {Zhang}, \citenamefont {Usechak},\ and\ \citenamefont
  {Lin}}]{Jiang:14}%
  \BibitemOpen
  \bibfield  {author} {\bibinfo {author} {\bibfnamefont {W.~C.}\ \bibnamefont
  {Jiang}}, \bibinfo {author} {\bibfnamefont {J.}~\bibnamefont {Zhang}},
  \bibinfo {author} {\bibfnamefont {N.~G.}\ \bibnamefont {Usechak}}, \ and\
  \bibinfo {author} {\bibfnamefont {Q.}~\bibnamefont {Lin}},\ }\href {\doibase
  http://dx.doi.org/10.1063/1.4890986} {\bibfield  {journal} {\bibinfo
  {journal} {Appl. Phys. Lett.}\ }\textbf {\bibinfo {volume} {105}},\
  \bibinfo {eid} {031112} (\bibinfo {year} {2014})}\BibitemShut {NoStop}%
\bibitem [{\citenamefont {Moss}\ \emph {et~al.}(2013)\citenamefont {Moss},
  \citenamefont {Morandotti}, \citenamefont {Gaeta},\ and\ \citenamefont
  {Lipson}}]{Moss:13}%
  \BibitemOpen
  \bibfield  {author} {\bibinfo {author} {\bibfnamefont {D.~J.}\ \bibnamefont
  {Moss}}, \bibinfo {author} {\bibfnamefont {R.}~\bibnamefont {Morandotti}},
  \bibinfo {author} {\bibfnamefont {A.~L.}\ \bibnamefont {Gaeta}}, \ and\
  \bibinfo {author} {\bibfnamefont {M.}~\bibnamefont {Lipson}},\ }\href
  {http://dx.doi.org/10.1038/nphoton.2013.183} {\bibfield  {journal} {\bibinfo
  {journal} {Nat Photon}\ }\textbf {\bibinfo {volume} {7}},\ \bibinfo {pages}
  {597} (\bibinfo {year} {2013})}\BibitemShut {NoStop}%
\bibitem [{\citenamefont {Rothman\emph{ et al}}(2013)}]{Rothman:13}%
  \BibitemOpen
  \bibfield  {author} {\bibinfo {author} {\bibfnamefont {L.~S.}\ \bibnamefont
  {Rothman\emph{ et al}}},\ }\href {\doibase
  http://dx.doi.org/10.1016/j.jqsrt.2013.07.002} {\bibfield  {journal}
  {\bibinfo  {journal} {J. Quant. Spectrosc. Radiat. Trans.}\ }\textbf {\bibinfo {volume} {130}},\ \bibinfo {pages} {4 }
  (\bibinfo {year} {2013})}\BibitemShut {NoStop}%
\bibitem [{\citenamefont {Preston}(1996)}]{Preston:96}%
  \BibitemOpen
  \bibfield  {author} {\bibinfo {author} {\bibfnamefont {D.~W.}\ \bibnamefont
  {Preston}},\ }\href
  {\doibase 10.1119/1.18457} {\bibfield  {journal} {\bibinfo  {journal}
  {Am. J. Phys.}\ }\textbf {\bibinfo {volume} {64}},\ \bibinfo
  {pages} {1432} (\bibinfo {year} {1996})}\BibitemShut {NoStop}%
\bibitem [{\citenamefont {Gorodetsky}\ \emph {et~al.}(2000)\citenamefont
  {Gorodetsky}, \citenamefont {Pryamikov},\ and\ \citenamefont
  {Ilchenko}}]{Gorodetsky:00}%
  \BibitemOpen
  \bibfield  {author} {\bibinfo {author} {\bibfnamefont {M.~L.}\ \bibnamefont
  {Gorodetsky}}, \bibinfo {author} {\bibfnamefont {A.~D.}\ \bibnamefont
  {Pryamikov}}, \ and\ \bibinfo {author} {\bibfnamefont {V.~S.}\ \bibnamefont
  {Ilchenko}},\ }\href
  {\doibase 10.1364/JOSAB.17.001051}{\bibfield  {journal} {\bibinfo  {journal} {J.
  Opt. Soc. Am. B}\ }\textbf {\bibinfo {volume} {17}},\ \bibinfo {pages} {1051}
  (\bibinfo {year} {2000})}\BibitemShut {NoStop}%
\bibitem [{\citenamefont {Cai}\ \emph {et~al.}(2000)\citenamefont {Cai},
  \citenamefont {Painter},\ and\ \citenamefont {Vahala}}]{Cai:00}%
  \BibitemOpen
  \bibfield  {author} {\bibinfo {author} {\bibfnamefont {M.}~\bibnamefont
  {Cai}}, \bibinfo {author} {\bibfnamefont {O.}~\bibnamefont {Painter}}, \ and\
  \bibinfo {author} {\bibfnamefont {K.~J.}\ \bibnamefont {Vahala}},\ }\href
  {\doibase 10.1103/PhysRevLett.85.74} {\bibfield  {journal} {\bibinfo
  {journal} {Phys. Rev. Lett.}\ }\textbf {\bibinfo {volume} {85}},\ \bibinfo
  {pages} {74} (\bibinfo {year} {2000})}\BibitemShut {NoStop}%
\end{thebibliography}

%


\end{document}